\documentclass[prl,twocolumn,showpacs,preprintnumbers,amsmath,amssymb,a4paper]{revtex4}
\usepackage{dcolumn}
\usepackage{bm}
\usepackage[dvips]{graphics}
\usepackage{natbib}
\begin{document}

\title{Equation of state for Entanglement in a Fermi gas}

\author{Christian Lunkes$^1$  }
\email{christian.lunkes@ic.ac.uk}
\author{ \v{C}aslav Brukner$^{1,2}$ }
\email{ c.brukner@ic.ac.uk}
\author{ Vlatko Vedral$^{3,4}$ }
\email{ v.vedral@ic.ac.uk} \affiliation{
$^1$QOLS, Blackett Laboratory, Imperial College London, London SW7 2BZ, England \\
$^2$Institut f\"{u}r Experimentalphysik, Universit\"{a}t Wien, Boltzmanngasse 5, A-1090 Vienna , Austria\\
$^3$The Schr\"odinger Institute for Mathematical Physics, Boltzmanngasse 9, Vienna,
Austria\\
$^4$The School of Physics and Astronomy, University of Leeds, Leeds LS2 9JT, England }

\date{\today}

\begin{abstract}
Entanglement distance is the maximal separation between  two entangled electrons in a
degenerate electron gas. Beyond  that distance, all entanglement disappears. We relate
entanglement distance to degeneracy pressure both for extreme relativistic and
non-relativistic systems, and estimate the entanglement distance in a  white dwarf.
Treating entanglement as a thermodynamical quantity, we relate the entropy of
formation and concurrence to relative electron distance, pressure, and temperature, to
form a new \textit{equation of state } for entanglement.
\end{abstract}


\maketitle                           

The relation between thermodynamics and entanglement is of fundamental importance, and
has been studied by many authors in the recent years
\cite{label1,label2,label3,label4,label5,label6}. The second law of thermodynamics
states that entropy $S$ can be regarded as a measure  of disorder. It is a macroscopic
variable and can be related to other state variables like pressure and temperature. In
quantum mechanics, the entropy of formation is  a measure of entanglement
\cite{label7}. Entanglement is a purely quantum mechanical feature and the entropy of
formation quantifies how much entanglement is present in a physical  system.
Entanglement, which is usually associated with the microscopic world, has recently
been shown to also become relevant on macroscopic scales \cite{label6}. Brukner and
Vedral \cite{label8} showed that thermodynamical quantities can also serve as
entanglement witnesses. All of this suggests that one may treat entanglement itself as
a thermodynamical quantity, and relate it to other macroscopic variables. This is the
principal motivation of the present work.
\newline
\newline
The main system studied in this letter is a Fermi gas, i.e. a system of noninteracting
fermions. Our treatment can be extended to other systems, e.g. interacting electrons
in a superconductor.
 Entanglement at zero temperature for noninteracting fermions has already been studied \cite{label9,label10}.
It was found that all entanglement vanishes if the relative distance $r$ between
electrons is greater then the entanglement distance $r_e\propto 1/k_F$, where   $k_F$
is the Fermi momentum.
 In this case, entanglement is purely due to
 particle statistics and not due to any physical interaction. At zero temperature, the electron gas is in its lowest energy configuration.
All energy states are occupied up to the Fermi energy $\epsilon_{F}$.  The only
pressure the system  exerts is the degeneracy pressure $P(n)$ (due to the Pauli
exclusion principle), which solely depends on its density $n$. We now want to relate
this pressure to entanglement. For the non-relativistic case, the expression for
pressure remains valid if the temperature is low compared to the Fermi temperature
$T_F=\epsilon_{F}/k_B$, where $k_B$ is the Boltzmann constant. This condition is met
if the density is sufficiently high, at which the fermion gas behaves as an ideal gas.
Since the particles in an ideal gas are not interacting, we can use the same treatment
as in \cite{label9,label10}, provided the density of the fermion gas is high enough.
For the extreme relativistic case, the expression for pressure is valid for all
temperatures.
\newline
\newline
We will first  relate $r_e$ to the degeneracy pressure $P$.
 We then estimate $r_e$ in white dwarfs. Third, we express
entropy of formation and concurrence, which  are measures of entanglement, as
functions of  degeneracy pressure,
 relative distance between the electrons $r$,  and temperature $T$.
 This represents our equation of state for entanglement.
 \newline
\newline
Vedral \cite{label9}, and  then Oh and Kim \cite{label10},
 first described entanglement of noninteracting electron gases. They used Green's function approach to
 find the subsystem  of the  two electrons,
 separated by a relative distance $r$. It is given by the  two-spin density matrix \cite{label10}:
\begin{equation}\label{eq:dm}
\rho_{12}(r)=\frac{n^2}{8}
[\delta_{\sigma_1,\sigma'_1}\delta_{\sigma_2,\sigma'_2}-f(r)^2\delta_{\sigma_1,\sigma'_2}\delta_{\sigma'_1,\sigma_2}
]\,,
\end{equation}
where $\sigma_1,\sigma_2,\sigma_1',\sigma_2'$ are  the spin variables
 of the two electrons under consideration. Following \cite{label9,label10}, the
state  is entangled iff the Peres-Horodecki (partial transposition)
 condition \cite{label11}  is satisfied :
\begin{equation}
f(r)^2>\frac{1}{2}\,,
\end{equation}
where $f$ at zero temperature and non zero temperature is given by:
\begin{equation}
f(r,0)=\frac{3}{rk_F^3}\int_0^{k_F} k\, \sin kr\, dk
\end{equation}
and
\begin{equation}\label{eq:ff}
f(r,T)=\frac{3}{rk_F^3}\int_0^{\infty} k  \, n_k \,\sin kr \,dk\,,
\end{equation}
where
\begin{equation}
  n_k=\frac{1}{exp(\frac{\epsilon_k -\mu}{k_B T}) +1}
\end{equation}
is the Fermi-Dirac distribution, $\mu$ is the chemical potential, and $\epsilon_k$ are
the energy levels. The region of entanglement is:
\begin{equation}\label{eq:in}
0\leq r < r_e \,,
\end{equation}
where $r$ is the relative distance between the two electrons, and the entanglement
distance $r_e$ is found to be inversely proportional to the Fermi momentum $k_F$:
\begin{equation}\label{eq:ot}
r_e=\frac{\zeta}{ k_F}\,.
\end{equation}
The proportionality constant $\zeta$ can be found numerically by solving
$f^2=\frac{1}{2}$. The integral in Eq.(\ref{eq:ff}) can be approximated for different
temperatures \cite{label10}. In general, $\zeta$ is a function of temperature, but for
$T$ sufficiently close to zero, $\zeta$ and the entanglement distance $r_e$ only
change slightly from the ones at zero temperature. However, the functional dependence
of the relation $k_F=\frac{\zeta}{r_e }$ remains the same for a gas with constant
density. A typical value for small temperatures is $\zeta \simeq 1.8$. We note that
there will be higher order entanglement present (3, 4 etc. electrons), but we will not
consider it here.
\newline
\newline
Eq.(\ref{eq:in}) is the upper bound for the distance of two entangled electrons. The
 entanglement distance $r_e$ is of the order of the Fermi wavelength, which
corresponds to the spread of the momentum wavefunction of the electrons that are close
to the Fermi sphere. In general, only these electrons  are of physical significance.
Eq.(\ref{eq:in}) is the most restrictive bound for entanglement distance and is valid
for all electrons in the gas. Any randomly chosen electrons that are closer then $r_e$
are definitely entangled. However, due to the spread in their wavefunction, two
electrons whose momenta lie far below the Fermi surface can also be entangled even if
they are further apart then $r_e$. This however implies the knowledge of their momenta
$k$, which is not accessible in this case. Eq.(\ref{eq:in}) was derived by making no
preselection of momenta. The function $f(r)$ in Eq.(\ref{eq:dm}) is computed by
summing over all possible momenta, from $k=0$ to $k=k_F$, for $T=0$. In order to have
entanglement, $f(r)$ must then satisfy the Peres-Horodecki criterion. In
\cite{label10}, the entanglement was already discussed as a function of the relative
distance between the electrons. It was found that the entanglement is maximal if the
relative distance between the electrons is zero. In this case, the spatial
wavefunctions fully overlap, and because of the antisymmetric nature of the two
electron wavefunction, the particles must have opposite spin. No spin configuration
($|\uparrow\downarrow\rangle, |\downarrow\uparrow\rangle$) is discriminated and
therefore the  two electrons are in the  maximally entangled Bell state,
$|\Psi^{-}\rangle=\frac{1}{\sqrt{2}}(|\uparrow\downarrow\rangle-|\downarrow\uparrow\rangle)$.
\newline
\newline
We now relate this entanglement to $P$. First we discuss the domain of validity of our
approach. Let us start by considering an electron  gas at $T=0$. All the electrons
 fill up the quantum states in accordance with
 the Pauli exclusion principle so that the total energy has its
 smallest value. The electrons have momenta from $p=0$
 to $p_F=\hbar k_F=(3\pi^2 n)^{\frac{1}{3}}\hbar$.
The radius of the Fermi sphere in momentum space
 is the Fermi momentum $k_F$, and $n=N/V$ is the density.
 The number of states in a volume element of width $dp$ is
 proportional to $p^2$. The total energy can then be obtained
by multiplying the number of states by $p^2/2m$ and integrating over all momenta. This
leads to the general equation  of state of a Fermi gas, from which the degeneracy
pressure can be deduced. We distinguish between the non-relativistic and the extreme
relativistic case \cite{label12}:
\begin{equation}
\label{eq:jk} P =
\begin{cases}
\frac{(3\pi^2)^\frac{2}{3} \hbar^2}{5 m} n^\frac{5}{3} & \text{non-rel.} \\\\
\frac{(3\pi^2)^\frac{1}{3}\hbar c}{4}   n^\frac{4}{3} & \text{rel.}\,,
\end{cases}
\end{equation}
where $m$ is the mass of the electron. The non-relativistic expression is  valid for
low temperatures, provided that $T \ll T_F$, which is the the Fermi temperature.
Rewriting this gives $T \ll \frac{(3\pi^2 \hbar^3 )^{\frac{2}{3}}}{2 m
k_B}n^{\frac{2}{3}}$, which is satisfied if the gas density is sufficiently high. A
degenerate electron gas has the peculiar property that it increasingly approaches the
ideal gas as the density increases \cite{label12}. This can be seen from the following
argument. Following Landau and Lifshitz \cite{label12}, we consider a plasma of
electrons and positively charged nuclei. When the electrons move independently as in
an ideal gas, the mean kinetic energy (which in order of magnitude is equal to the
Fermi energy) is much higher than the Coulomb interaction. For a nucleus of charge
$Ze$ and with the mean distance between the electrons and the nuclei being $a$:
\begin{align*}
&E_{Coulomb}\ll E_{kinetic}\\
&\frac{Ze^2 }{a}\ll \frac{{p_F}^2}{2m} \\&\frac{Ze^2
n^\frac{1}{3}}{Z^\frac{1}{3}}\ll\frac{{ \hbar}^2n^\frac{2}{3}}{2m}\quad with \quad
a\sim
(\frac{Z}{n})^\frac{1}{3}\\
&n\gg(\frac{e^2m}{\hbar^2})^3Z^2 \,.
\end{align*}
There is no electron interaction if the gas density is sufficiently high.
\newline
\newline
Let us now relate $r_e$ to the degeneracy pressure.
 The non-relativistic expression for the degeneracy pressure
 is given by Eq.(\ref{eq:jk}), which is
 valid at low temperatures if the density is high. High density means no
 electron interaction, so we  use the same treatment as
 in  \cite{label9,label10}. The extreme relativistic expression for the degeneracy pressure
 is valid at all temperatures. The Fermi momentum can
 be expressed in terms of the  density $n$, and
Eq.(\ref{eq:ot}) is rewritten as:
\begin{equation}\label{eq:ab}
r_e=\frac{\zeta}{(3\pi^2 )^{\frac{1}{3}}n^{\frac{1}{3}}}\,.
\end{equation}
The entanglement distance is approximately equal to the average distance between the
electrons in the gas. Rewriting the degeneracy pressure in terms of $r_e$, we obtain
for the non-relativistic and  extreme relativistic case:
\begin{equation}\label{eq:dp}  P =
\begin{cases}
 \frac{ \zeta^5\hbar^2}{15 \pi^2 m } \: r_e^{-5} & \text{non-rel.}  \\\\
 \frac{{\zeta'}^4\hbar c}{12 \pi^2  } \: r_e^{-4}  & \text{rel.} \,.
\end{cases}
\end{equation}
This expression is valid if $T \ll T_F$. The constants $\zeta$ and $\zeta'$ are
different because in the extreme relativistic limit, the density is much higher then
in the non-relativistic case. When the degeneracy pressure of a given gas is high,
$r_e$ is small and the Fermi momentum is high. This means that the electrons are
moving at high velocities and are localized in space with no overlap of their
wavefunctions, hence no entanglement is present in this limit.
\newline
\newline
We now give an estimate for $r_e$ in white dwarfs.
 White dwarfs are examples of highly degenerate matter.
They are objects of extremely high density. We express the electron density as:
\begin{equation}\label{eq:op}
n=\frac{Z \rho}{A m_H}\,,
\end{equation}
where $Z$ and $A$ are the number of protons and nucleons and $m_H$ is the mass of the
Hydrogen atom. If we now assume that the density is constant, we can deduce
entanglement distance for a white dwarf of mass $M$ and radius $R$. The density is:
\begin{equation}
\rho=\frac{M}{4/3\pi R^3}\,.
\end{equation}
Using Eq.(\ref{eq:op}),  we write:
\begin{equation}\label{eq:ul}
r_e\propto R M^{-\frac{1}{3}}\,.
\end{equation}
We estimate the entanglement distance of
  SIRIUS B, which is  a carbon-oxygen white dwarf, of mass
$1 M_\odot$,  radius  $R=0.008 R_\odot$, and a temperature of $T=27000 K$
\cite{label13}. Assuming non-relativistic electrons, and calculating the Fermi
temperature gives $T/T_F \sim 10^{-6}$, which justifies our low temperature
approximation. From Eq.(\ref{eq:ul}), the  entanglement distance is found to be $r_e
\simeq 6 \times 10^{-13} m $. This is two orders of magnitude bigger then the size of
a typical nucleus. Note that here entanglement length is much smaller than the size
$R$ of the system. This should be contrasted with the limit of entanglement length
being of the order of $R$, i.e $r_e \simeq R$, for which the mass of the system is
 $m\simeq 2.7 \times 10^{-27} kg$, which is about the mass of a nucleus.
\newline
\newline
Let us finally express the entropy of formation  as a function of the degeneracy
pressure,
 relative distance between the electrons $r$,  and temperature $T$.
Relating the Fermi momentum to the degeneracy pressure for the non-relativistic and
extreme relativistic case:
\begin{equation}  k_F =
\begin{cases}
   \left(%
\frac{ 15\pi^2m}{\hbar^2}
\right)^{\frac{1}{5}}P^\frac{1}{5}       & \text{non-rel.}  \\\\
    \left(%
\frac{ 12\pi^2}{\hbar c}
\right)^{\frac{1}{4}}P^\frac{1}{4}       & \text{rel.}\,,
\end{cases}
\end{equation}
so Eq.(\ref{eq:ff}) becomes
\begin{equation} \label{eq:as} f_{r,P,T} =
\begin{cases}
  \frac{\gamma}{r P^\frac{3}{5}}\int_0^{\infty} k  \, n_k \,\sin kr \,dk     & \text{non-rel.}
  \\\\
  \frac{\gamma'}{r P^\frac{3}{4}}\int_0^{\infty} k  \, n_k \,\sin kr \,dk    & \text{rel.}
  \,,
\end{cases}
\end{equation}
with
\begin{equation}
\gamma= \left(%
\frac{\hbar^2 3^\frac{5}{3}}{ 15\pi^2m}
\right)^{\frac{3}{5}} \quad  \text{and}\quad
\gamma'= \left(%
\frac{\hbar c 3^\frac{4}{3}}{ 12\pi^2}
\right)^{\frac{3}{4}}.
\end{equation}
 We now  describe the system as it is
compressed. If the density $n$  changes, so does the Fermi momentum $k_F=(3\pi^2
n)^{\frac{1}{3}}$ and the degeneracy pressure $P$. This corresponds to a white dwarf,
where the particles are forced into a smaller and smaller volume by the gravitational
pressure. This process can continue until the extreme relativistic limit is reached
where the energy of the electrons becomes large compared to $mc^2$.

The integrals in Eq.(\ref{eq:as}) can be solved numerically for a given temperature.
At low temperatures,  we take the chemical potential to be $\mu\simeq\epsilon_{F}$.
All the information needed to compute entanglement is given in the two particle
density matrix Eq.(\ref{eq:dm}). From this density matrix, any entanglement measure
$E$ can be computed as a function of $f_{r,P,T}$. We express the entropy of formation
in terms of electron distance, degeneracy pressure and temperature  for an electron
gas. The \textit{equation of state for entanglement} becomes: {\small{
\begin{equation}\label{eq:qe}
E_{F}(r,P,T)=h\left[%
 \frac{1}{2}+ \frac{1}{2}\sqrt{1-\left(%
 \frac{2f_{r,P,T}^2-1}{2-f_{r,P,T}^2}%
\right)^2}%
\right]\,,
\end{equation}
}}with  $h(y)=-y\log_2 y-(1-y)\log_2 (1-y)$, and $f_{r,P,T}$ is either the
non-relativistic or extreme relativistic expression. We can also use the concurrence
\cite{label14} to express  entanglement:
\begin{equation}\label{eq:ue}
C(r,P,T)=max\{\frac{2f_{r,P,T}^2-1}{2-f_{r,P,T}^2},0\}\,.
\end{equation}
 Consider a non-relativistic gas with high density  at zero temperature, which undergoes
 compression. This is shown in Fig. 1. If we fix  the relative distance $r$ of the two electrons  to be
smaller then the entanglement distance $r_e$, then the two electrons are certainly
entangled. The smaller the distance $r$ is, the more entangled are the electrons. If
we now compress the system,  the density $n$ increases, and  $r_e$ in Eq.(\ref{eq:ab})
decreases. The entanglement distance $r_e$ approaches the relative electron distance
$r$ from above, and the entanglement between the two electrons decreases. If the
system is compressed further, $r_e$ becomes smaller then $r$, and entanglement
vanishes. This means that the  electrons are forced into a smaller and smaller volume,
so their momenta increase and their wavefunctions overlap less, until their is no
overlap and therefore no entanglement.
\newline
\newline
Eq.(\ref{eq:qe}) or Eq.(\ref{eq:ue}) are not equations of state in the "classical
sense" of that word. We call them like that
 because they treat entanglement on the same footing as pressure and temperature,
  which are classical thermodynamical variables.
 The only microscopic variable left is the relative distance between the electrons.
 \begin{figure}
\begin{center}
\resizebox{8cm}{8cm}{\includegraphics{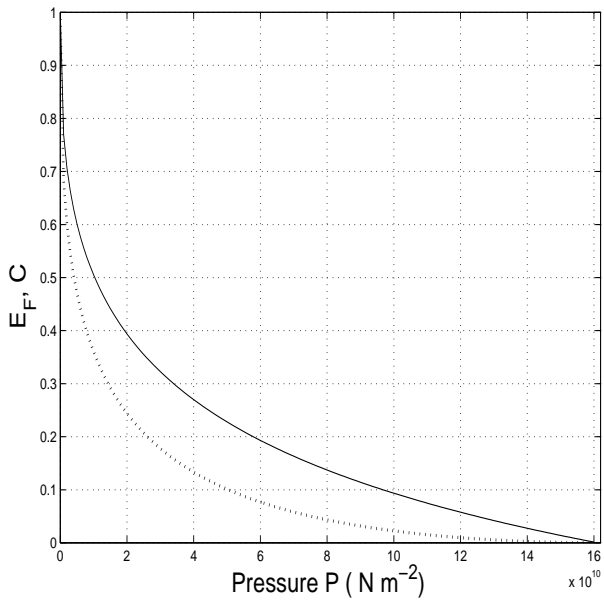}}
\end{center}
\caption{Concurrence $C$ (solid line) and Entropy of formation $E_F$ (dotted line) at
zero temperature with fixed $r=10^{-10}m $ as a function of non-relativistic
degeneracy pressure $P$, for a gas with initial entanglement distance $r_e=10^{-8} m
$.}
\end{figure}
If we do not want a microscopic variable to appear in the equation of state, we can
  average over the region where entanglement between two electrons certainly
 exists. For convenience, first define the variable $x=k_Fr$. The average
then implies integrating from $x=0$ to $x=x(T)$. We find $x(T)$ by solving
$f(x,T)^{2}=\frac{1}{2}$. We can now find the average amount of entanglement for
electron distances where there exists entanglement, using a general integrable
 entanglement measure $E$:
\begin{equation}
\langle E(T)\rangle=\frac{1}{x(T)}\int_{x=0}^{x=x(T)} E(x,T) dx\,.
\end{equation}
\newline
\newline
In  conclusion, we have considered a degenerate electron gas with high density at  low
temperatures, and an extremely relativistic electron gas. We first related
entanglement distance to degeneracy pressure. We then estimated entanglement distance
for  the simplest model of a white dwarf. For the first time, the entropy  of
formation and the concurrence  was related to temperature, pressure and relative
distance between the electrons. It is valid for temperatures which are sufficiently
close to zero and high density for the non-relativistic case, and for all temperatures
for the extreme relativistic case. In this work entanglement
 is purely due to particle statistics (Pauli exclusion principle) and not to fermion interaction. Future work
would consist of investigating entanglement for  fermions that physically interact. It
would also be interesting to extend the analysis to  multipartite entanglement and to
see if one could indeed treat entanglement as a thermodynamical variable, and write an
equation for entanglement similar to the the fundamental thermodynamic equation for
energy $U$, $dU = T dS-P dV$.

We are grateful to Marcelo Fran\c{c}a Santos for helpful discussions.

\end{document}